\documentclass[%
 reprint,
 amsmath,amssymb,
 aps,
]{revtex4-2}

\usepackage{graphicx}
\usepackage{dcolumn}
\usepackage{bm}
\usepackage{hyperref}

\begin{document}


\title{The cosmic coincidences of primordial-black-hole dark matter}

\author{Yi-Peng Wu$^{a}$}
\author{Elena Pinetti$^{abf}$}
\author{Joseph Silk$^{cde}$}


\affiliation{$^a$Laboratoire de Physique Th\'{e}orique et Hautes Energies (LPTHE), \\
	UMR 7589 CNRS and Sorbonne Universit\'{e}, 4 Place Jussieu, F-75252, Paris, France}
\affiliation{$^b$Dipartimento di Fisica, Universit\'{a} di Torino and INFN, Sezione di Torino, via P. Giuria 1, I-10125 Torino, Italy}
\affiliation{$^c$Institut d’Astrophysique de Paris, UMR 7095 CNRS and Sorbonne Universit\'{e}, 98 bis boulevard Arago, F-75014 Paris, France}
\affiliation{$^d$Department of Physics and Astronomy, The Johns Hopkins University, 3400 North Charles Street, Baltimore, Maryland 21218, USA}
\affiliation{$^e$Beecroft Institute for Particle Astrophysics and Cosmology, University of Oxford, Keble	Road, Oxford OX1 3RH, United Kingdom}
\affiliation{$^f$Theoretical Astrophysics Department, Fermi National Accelerator Laboratory, Batavia, Illinois, 60510, USA}




\date{\today}

\begin{abstract} 
If primordial black holes (PBHs) contribute more than 10\% of the dark matter (DM) density, their energy density today is of the same order as that of the baryons. Such a cosmic coincidence might hint at a mutual origin for the formation scenario of PBHs and the baryon asymmetry of the Universe. Baryogenesis can be triggered by a sharp transition of the rolling rate of inflaton from slow-roll to (nearly) ultraslow-roll phases that produce large curvature perturbations for PBH formation in single-field inflationary models. We show that the baryogenesis requirement drives the PBH contribution to DM, along with the inferred PBH mass range, the resulting stochastic gravitational wave background frequency window, and the associated cosmic microwave background tensor-to-scalar ratio amplitude, into potentially observable regimes.
\end{abstract}

\maketitle

\section{Introduction}
Primordial black holes (PBHs) are one of the most interesting dark matter (DM) candidates that have been severely constrained by joint astrophysical and cosmological observations \cite{Carr:2009jm, Poulin:2016anj, Carr:2016drx, Clark:2016nst, Boudaud:2018hqb, Ballesteros:2019exr, DeRocco:2019fjq, Dasgupta:2019cae, Laha:2019ssq, Kim:2020ngi, Laha:2020ivk, Carr:2020gox,Cang:2020aoo,Cang:2021owu}.
The asteroid-mass window $M_{\rm PBH}/M_\odot \sim 10^{-16} - 10^{-12} $ for PBHs to be all DM \cite{Montero-Camacho:2019jte,Smyth:2019whb,Niikura:2017zjd,Niikura:2019kqi} will be tested in the near future by femtolensing of gamma ray bursts \cite{Nemiroff:1995ak, Katz:2018zrn, Jung:2019fcs}, microlensing of x-ray pulsars \cite{Bai:2018bej}, primary photons measured by next-generation MeV detectors \cite{Ballesteros:2019exr, Coogan:2020tuf}, neutron star disruption \cite{Capela:2013yf}, white dwarf explosions \cite{Graham:2015apa}, MeV photons from the Galactic Center \cite{Ray:2021mxu} and radio emission measurable with the next-generation radio telescopes \cite{Dutta:2020lqc}. Even if PBHs only occupy a small fraction of the DM density today, their existence could be (or could have been \cite{Franciolini:2021tla}) probed by gravitational wave observations through   binary mergers, either mutual   or with neutron stars  (for recent reviews, see \cite{Carr:2020xqk,Green:2020jor, Carr:2020xqk}). 

If PBHs really constitute more than 10\% of the DM density today, the PBH density up to matter-radiation equality is of the same order as that of the baryons, namely $\Omega_{\rm PBHeq}/\Omega_{\rm Beq}\sim \mathcal{O}(1)$. Such a cosmic coincidence would hint at a mutual origin for PBHs and the baryon asymmetry of the Universe. In this Letter, we argue that the PBH-baryon density coincidence could be a natural consequence due to baryogenesis triggered by inflation models for PBH formation. We show that the abundance of PBHs and baryons are indirectly correlated to each other via the dynamics of inflation, and, thus, the scenario is very different from  previous suggestions that existing PBHs create baryon asymmetry \cite{Turner:1979bt,Barrow:1990he,Majumdar:1995yr,Baumann:2007yr,Bambi:2008hp,Hamada:2016jnq,Hooper:2020otu,Ambrosone:2021lsx,JyotiDas:2021shi,Datta:2020bht} or account for cosmic coincidence \cite{Fujita:2014hha,Garcia-Bellido:2019vlf,Carr:2019hud}.
(See, also, \cite{Flores:2020drq} for cosmic coincidence from asymmetric DM \cite{Bell:2011tn,vonHarling:2012yn,Petraki:2013wwa} collapses into PBHs.)

\section{\label{sec:USR inflation} PBHs from Inflation}
Let us focus on single-field inflation for PBH formation \cite{Cicoli:2018asa,Yokoyama:1998pt,Motohashi:2019rhu,Saito:2008em,Germani:2017bcs,Garcia-Bellido:2017mdw,Motohashi:2017kbs,Cheng:2018qof,Liu:2020oqe,Biagetti:2018pjj,Ballesteros:2020qam,Byrnes:2018txb,Bhaumik:2019tvl,Xu:2019bdp,Atal:2018neu,Kannike:2017bxn,Ragavendra:2020sop,Ozsoy:2018flq,Taoso:2021uvl,Saito:2008em,Ozsoy:2019lyy,Ballesteros:2017fsr,Mishra:2019pzq}.
The generic assumption is that the inflaton $\phi$ receives a sudden deceleration on comoving scales $k_0 \sim 10^{12} - 10^{15}$ Mpc$^{-1}$ which leads to a sharp decrease of the first slow-roll parameter $\epsilon_H \equiv -\dot{H}/H^2$ and, thus, largely enhances the power spectrum of the curvature perturbation $P_\zeta$ \cite{Leach:2001zf}. Such an enhancement is due to the temporal dominance of the entropy mode in the curvature perturbation \cite{Ragavendra:2020sop,Leach:2000yw,Ng:2021hll,Drees:2019xpp}.

 The key parameter for realizing the enhancement of $P_\zeta$ is the rate of rolling:
\begin{equation}\label{USR_delta_definition}
\delta = \frac{\ddot{\phi}}{H\dot{\phi}} < -3/2, \quad t_{0} < t < t_{\ast},
\end{equation}
where $\delta \rightarrow 0$ is the standard slow-roll inflation and $\delta\rightarrow -3$ is the so-called ultraslow-roll (USR) limit \cite{Motohashi:2014ppa,Martin:2012pe,Anguelova:2017djf,Tsamis:2003px,Kinney:2005vj}. 
 With $\delta < -3$, $P_\zeta$ can exhibit a spiky peak at a desired scale for producing a nearly monochromatic distribution of PBH mass.

We consider that $\delta$ takes constant values in different phases of inflation. 
In terms of the $e$-folding numbers, $N\equiv \ln a$, where $k_{0}/k_{\rm \ast} \approx a(t_0)/a(t_\ast) = e^{N_{\ast}-N_0}$ with $N_0 \equiv 0$,
the duration $N_{\ast}$ (for $\delta$ having a negative value) is the fundamental parameter that controls the spectral amplitude of $P_\zeta$ (and, thus, the PBH abundance).

The analytic structure of the USR power spectrum at the end of inflation ($N = N_{\rm end}$) has been intensively investigated \cite{Byrnes:2018txb,Liu:2020oqe,Cheng:2018qof,Ozsoy:2019lyy,Ng:2021hll,Ballesteros:2020sre}. Dilatation symmetry of the de Sitter background requires the momentum scaling at each phase with different values of $\delta$ to satisfy \cite{Liu:2020oqe,Ng:2021hll}
\begin{eqnarray}\label{CR_template}
P_\zeta = \left\{
\begin{array}{ll}
A_{\rm CMB} & k < k_{\rm min}, \\
A_{\rm PBH} (k/k_0)^4, & k_{\rm min} < k < k_0, \\
A_{\rm PBH} (k/k_0)^{6 + 2\delta}, & k_0 < k < k_{\rm end},
\end{array}
\right.
\end{eqnarray}
where $A_{\rm CMB}  \approx 2.2\times 10^{-9}$ measured on comis microwave background (CMB) scales has negligible contribution to PBH formation. 
$k_0$ is the pivot scale for the enhancement and shall be fixed by the desired peak scale $M_{\rm PBH}$ in the PBH mass function.
$k_{\rm min} \approx k_0 (A_{\rm CMB}/A_{\rm PBH})^{1/4}$ is the beginning scale of the $k^4$ growth driven by the Leach-Sasaki-Wands-Liddle mechanism \cite{Leach:2001zf} (sometimes also called the steepest growth \cite{Byrnes:2018txb,Carrilho:2019oqg}).
The amplitude $A_{\rm PBH}$ is determined by USR parameters as
\begin{equation}\label{A_zeta_analytic}
A_{\rm PBH} \approx  A_{\rm CMB} \left(\frac{k_0}{k_{\ast}} \right)^{6 + 4 \delta}
= A_{\rm CMB} e^{-N_{\ast}(6+4\delta)}. 
\end{equation}
Note that in the template \eqref{CR_template} one should use the value of $\delta < -3$ found in the deceleration phase $N_0 < N < N_\ast$, since \eqref{USR_delta_definition} must become positive for $N > N_\ast$ to increase $\epsilon_H$ and terminate inflation. The positive rolling rate in the final acceleration phase ($N > N_\ast$) is constrained by $\delta$ in the deceleration phase (with respect to the conformal symmetry due to nonviolation of the adiabatic condition \cite{Ng:2021hll}) so that the scaling of $P_\zeta (k)$ for $k_\ast < k < k_{\rm end}$ is the same as $k_0 < k < k_\ast$.



In the case of exact USR ($\delta \rightarrow -3$), the inflaton potential $V(\phi)$ is completely flat so that $\phi$ is exactly massless, where quantum diffusion led by short wavelength modes well inside the horizon may have an important impact on the classical trajectory of $\phi$ \cite{Ballesteros:2017fsr,Biagetti:2018pjj,Pattison:2021oen}. A non-Gaussian tail in the high-sigma limit of the probability distribution of $\zeta$ can significantly raise the resulting PBH abundance from USR inflation \cite{Biagetti:2018pjj,Ezquiaga:2019ftu,Pattison:2021oen,Figueroa:2020jkf,Biagetti:2021eep}, indicating the real amplitude $A_{\rm PBH}$ estimated by the Gaussian spectrum \eqref{CR_template} (based on the linear relation $\zeta = -H/\dot{\phi} \delta\phi$) should be smaller than expected. To suppress the effect of quantum diffusion, we adopt an upper bound $\delta < -3.1$, which corresponds to an effective mass $m_{\phi} \equiv (V_{\phi\phi})^{1/2} > H_\ast/2$ for the inflaton fluctuation $\delta\phi$ \cite{Wu:2021mwy}.

\section{\label{sec:baryogenesis} Baryogenesis via inflation}
Now, we show that baryogenesis can be triggered by USR inflation.
Scalar fields naturally develop large vacuum expectation values (VEVs) during inflation due to the high energy background expansion at the scale of $H_\ast$ (possibly as high as $10^{13 -14}$ GeV \cite{Tristram:2020wbi,Akrami:2018odb}). 
These large VEVs provide suitable initial conditions for baryogenesis driven by the Affleck-Dine (AD) mechanism \cite{Affleck:1984fy,Linde:1985gh,Dolgov:1991fr,Dine:2003ax,Dine:1995kz}:
(1) The stochastic nature of the inflationary fluctuations always allows CP-violating VEVs arising from theories with CP invariant Lagrangian \cite{Wu:2020pej,Dine:1995kz}, and (2) the post-inflationary relaxation of a $B$ or $B - L$ violating scalar condensate is an out-of-equilibrium process. 

A possible realization for the USR inflation to affect the dynamics of a charged scalar $\sigma$ is given by
\begin{align}\label{Lagrangian_AD}
\mathcal{L} =& \mathcal{L}_\phi +\vert \partial\sigma\vert^2 + m_\sigma^2 \vert \sigma\vert^2 
+\frac{c_1}{\Lambda} \left\vert  \sigma^2\right\vert \square \phi \\\nonumber
&+ \frac{c_2}{\Lambda}\partial_\mu \phi \left[\sigma\partial^\mu \sigma + \sigma^\ast\partial^\mu \sigma^\ast\right] + 
\mathcal{O}(\Lambda^{-2})\cdots,
\end{align} 
where $c_1$ and $c_2$ are real constants of $\mathcal{O}(1)$. CP invariance is imposed on the Lagrangian for demonstrative purposes, yet it is not a necessary condition for AD baryogenesis. Similar couplings for enhanced charged scalar production from rolling inflaton as a chemical potential can be found in \cite{Wang:2019gbi,Wang:2020ioa,Bodas:2020yho}. We ask $H_\ast \ll \Lambda \leq M_P$ for the cutoff $\Lambda$. Note that the $c_2$ term violates the conserved current $j^\mu = i(\sigma^\ast \partial_\mu\sigma - \sigma\partial_\mu\sigma^\ast)$, which is identified as a baryon number for convenience.

In terms of the mass eigenstates $\sigma_\pm$, where $\sigma \equiv (\sigma_{-} + i \sigma_{+})/\sqrt{2}$, the charged scalar is decomposed into a pair of decoupled canonical real scalars, $\mathcal{L}_{\sigma_\pm} = \frac{1}{2}(\partial\sigma_\pm)^2 + \frac{1}{2} m_\pm^2 \sigma_\pm^2$, with asymmetric (nondegenerate) masses as
\begin{align}\label{mass_pm}
m_\pm^2 = m_\sigma^2 + \frac{c_1 \pm c_2}{\Lambda} \square\phi .
\end{align}
One can see that the phase transition of $\delta$ for PBH formation also changes the effective masses as $\square\phi = -\ddot{\phi} -3H\dot{\phi} \approx - (\delta + 3)\sqrt{2\epsilon_H} M_PH_\ast^2$. 
For $m_\sigma \sim H_\ast$ and $\Lambda > 0.1 M_P$, the sharp decrease of $\epsilon_H$ for the $P_\zeta$ enhancement usually leads to $m_\pm \approx m_\sigma$ in the $\phi$-deceleration (USR) phase.

The sudden transition of $m_\pm$ from the primary slow-roll phase (with $\delta \sim 0$) to the deceleration phase with $\delta < -3$ drives the original VEVs of $\sigma_\pm$ out of equilibrium in their potential, triggering the coherent motion of these scalar condensates. The analytic solutions for the coherent motion of $\sigma_\pm$ are given in \cite{Wu:2021mwy}. At the end of inflation, the VEVs of the mass eigenstates are led by
\begin{align}\label{initial_condition_mass_eigenstates}
\sigma_\pm \sim e^{-\Delta_\pm^- N_{\rm end}}, \quad \dot{\sigma}_\pm \sim -\Delta_\pm^- e^{-\Delta_\pm^- N_{\rm end}},
\end{align} 
where $\Delta_\pm^- \equiv 3/2 - \sqrt{9/4 - m_\pm^2/H_\ast^2}$ is nothing but the negative branch of the conformal weight for a massive scalar in de Sitter \cite{Antoniadis:2011ib}. The late-time approximation used in \eqref{initial_condition_mass_eigenstates} applies when $m_\pm/H_\ast < 3/2$.

Assuming the standard reheating process driven by the coherent oscillation of $\phi$, one can numerically solve the relaxation of $\sigma_\pm$ from the end of inflation to reheating completion (or  radiation domination) \cite{Wu:2020pej,Wu:2021mwy}. Here, we consider the decay of $\phi$ into radiation via a perturbative channel with a decay width $\Gamma_\phi$. Thus, the approximated time scale at the beginning of radiation domination is $t_r \sim 1/\Gamma_\phi$.
The final baryon asymmetry in radiation domination reads
\begin{align}\label{Y_B__final}
Y_B = \frac{n_B(t_r)}{s(t_r)} =  \frac{\sigma_{+}(t_r)\dot{\sigma}_-(t_r) - \sigma_{-}(t_r)\dot{\sigma}_+ (t_r)}{s(t_r)} ,
\end{align}
where $s(t) \approx 2\pi^2g_\ast T^3(t)/45$ is the entropy production and $T = (\frac{90}{\pi^2 g_\ast} M_p^2 H^2 )^{1/4}$ is the temperature. 
Note that $Y_B = Y_B(\delta, N_\ast, N_{\rm end})$ as those parameters of USR inflation enter through the initial conditions \eqref{initial_condition_mass_eigenstates}.

\begin{figure}[t]
	\begin{center}
		\includegraphics[width=6 cm]{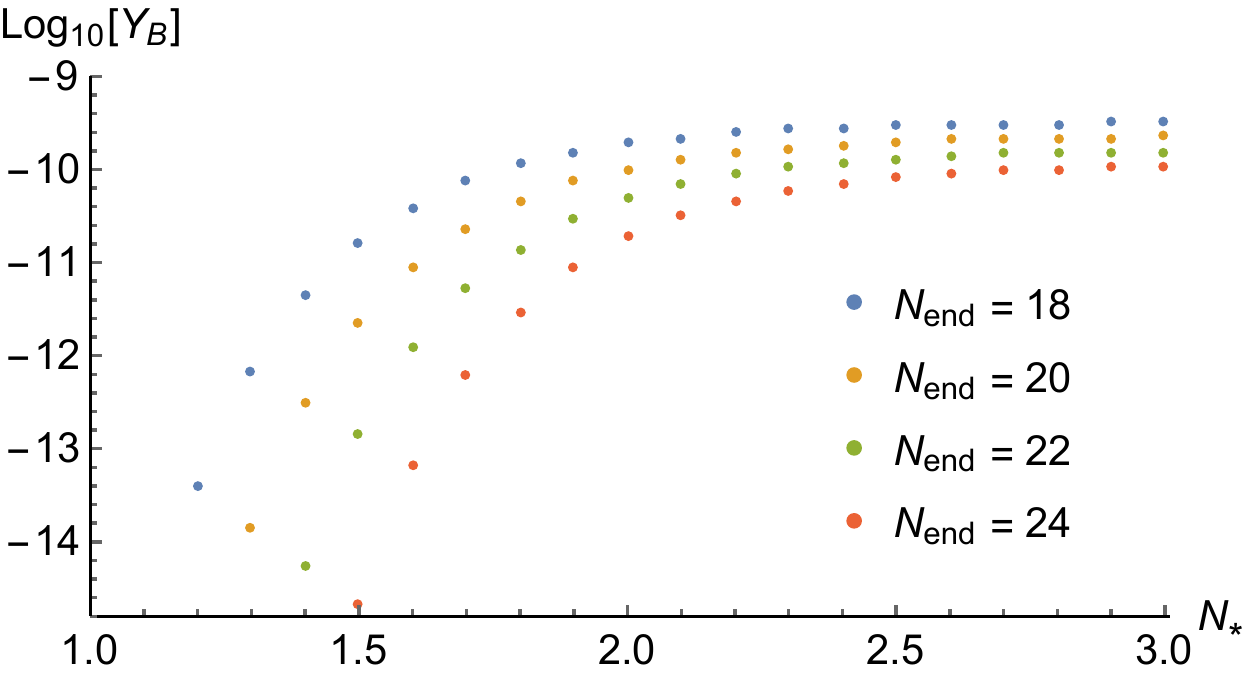}
	\end{center}
	\caption{\label{fig.YB_Ns} The final baryon asymmetry in radiation domination with $\delta = -3.15$, $m_\sigma = H_\ast/2$, $\Lambda = 0.3 M_P$, where $H_\ast = 2.37 \times 10^{13}$ GeV and $\Gamma_\phi = 10^{13}$ GeV are used.}
\end{figure}

We highlight the generic property of the final baryon asymmetry with examples given in Fig.~\ref{fig.YB_Ns}. In general, $Y_B$ is sensitive to $N_{\rm end}$ since initial conditions \eqref{initial_condition_mass_eigenstates} are exponentially diluted by the $e$-fold numbers, but it approaches a constant value when $N_\ast \sim \mathcal{O}(1)$ depending on the value of $\delta$. For $\delta = -3.15$, examples in Fig.~\ref{fig.YB_Ns} indicate that $Y_B \simeq$ const. when $N_\ast \gtrsim 2$, which corresponds to $A_{\rm PBH} \gtrsim 10^{-3}$. This asymptotic constant behavior of $Y_B$ is the most important property for resolving the coincidence problem for PBH DM.

\section{\label{sec:PBHDM}PBH dark matter}
As the inflaton decays during reheating, the enhanced curvature perturbation on scales $k > k_0$ is inherited by the density perturbation of the radiation.  
Soon after  reentry into the  horizon in radiation domination, PBHs are formed at high-sigma peaks of the density contrast $\Delta$ smoothed over a given comoving scale $R = 1/(a H) = 1/ k$. The comoving scale $R$ can be expressed in terms of the horizon mass parameter $M_H =  \frac{4\pi}{3}H^{-3} \rho_R$, with $\rho_R$ the energy density of the radiation dominated Universe, as
\begin{equation}
R(M_H) = \frac{1}{k_{\rm eq}} \left(\frac{M_H}{M_{\rm eq}}\right)^{1/2} \left(\frac{g_\ast}{g_{\rm eq}}\right)^{1/6},
\end{equation}
where $M_{\rm eq} = 2.9 \times 10^{17} M_\odot$ and $g_{\rm eq} \approx 3$ are the horizon mass and the number of relativistic degrees of freedom at matter-radiation equality. We use $k_{\rm eq} = 0.01 \textrm{Mpc}^{-1}$ and $g_\ast = 106.75$ for $M_H < 1.5 \times 10^{-7} M_\odot$ where the temperature of the Universe is higher than $300$ GeV. 

The mass fraction $\beta(M_{\rm PBH}, M_H)$ of a flat Universe that collapses into PBHs with  mass $M_{\rm PBH}$ at a given horizon mass $M_H$ can be obtained from the density parameter $ \Omega_{\rm PBH} (R)$ of PBHs at the corresponding scale $R(M_H)$ as $\beta(M_{\rm PBH}, M_H) = d\Omega_{\rm PBH} /d\ln M_{\rm PBH}$, where $\Omega_{\rm PBH}$ is usually estimated via  threshold statistics:
\begin{align}\label{threshold_statistics}
\Omega_{\rm PBH} (R) =& \int\cdots\int_{\Delta_c}^{\infty} \frac{M_{\rm PBH}}{M_H} \\\nonumber
&\times f_c(y_i) P(\Delta, y_i, \sigma_i) d\Delta dy_1\cdots dy_i.
\end{align}
Here $P(\Delta, y_i, \sigma_i)$ is the joint probability distribution of $\Delta$ and $y_i$ are components of its first and second order spatial derivatives. $f_c(y_i)$ describes spatial constraints to ensure the selected peaks are local maxima in space \cite{Bardeen:1985tr,Young:2014ana,Green:2004wb,Suyama:2019npc,Wu:2020ilx}.
All $y_i$ are Gaussian random fields if $\Delta$ is Gaussian. 
$\sigma_i$ stands for the $i$th spectral moment of the Gaussian field $\Delta$ smoothed by the window function $W(kR)$. The form of $\sigma_i$ is defined as
\begin{equation}\label{spectral_moment}
\sigma_i^2(R) = \int_{0}^{\infty} k^{2i} W^2(kR) P_{\Delta}(k) d\ln k,
\end{equation} 
where $P_{\Delta}$ is the dimensionless power spectrum.
$\Delta_c$ is the threshold value above which the density contrast will collapse to form a PBH.
Therefore, the mass fraction $\beta = \beta(M_H, \Delta_c, \sigma_i)$ is a general function of the smoothed scale $R$ (or, namely, $M_H$), the threshold $\Delta_c$, and the spectral moment $\sigma_i$. 


Even for inflation close to the USR limit ($\delta \rightarrow -3$), we find that $M_{\rm PBH}\approx M_H$ can be a good approximation for resolving the PBH mass function $f(M_{\rm PBH})$ \cite{Wu:2021mwy}. 
Such a monochromatic relation leads to a simple expression of the PBH density at matter-radiation equality 
\begin{equation}\label{PBH_density_eq}
\Omega_{\rm PBHeq} = \int \beta( M_H) \left(\frac{M_{\rm eq}}{M_H}\right)^{1/2} d\ln M_H,
\end{equation}  
where $(M_{\rm eq}/M_H)^{1/2}\sim a_{\rm eq}/a$ accounts for the relative growth of PBH density during radiation domination. The PBH mass function defined from the PBH-to-DM ratio,
\begin{equation}
f_{\rm PBH} \equiv \frac{\Omega_{\rm PBHeq}}{\Omega_{\rm DMeq}} = \frac{1}{\Omega_{\rm DMeq}} \int f(M_H) d\ln M_H,
\end{equation}
implies $f(M_H) = \beta( M_H) (M_{\rm eq}/M_H)^{1/2}/\Omega_{\rm DMeq}$.


\section{The cosmic coincidence}
In the standard $\Lambda$-cold dark matter (CDM) Universe \cite{Aghanim:2018eyx}, the cold dark matter density today $\Omega_{\rm CDM0} =0.265$ and the redshift $z_{\rm eq} = 3402$ gives $\Omega_{\rm CDMeq} = 0.42$ and $\Omega_{\rm Beq} = m_B n_{\rm Beq} = 0.08$. This shows that $\Omega_{\rm PBHeq}/\Omega_{\rm Beq} \simeq 0.5 - 5$ for $f_{\rm PBH} = 0.1 - 1$.
Here $m_B = 0.938$ GeV is the averaged nucleon mass and $n_{\rm Beq} = \vert Y_B\vert s(t_{\rm eq})$ is the baryon number density at matter-radiation equality. Using $H_{\rm eq} = H_0 \sqrt{\Omega_{\Lambda0} + 2 \Omega_{m0}(a_0/a_{\rm eq})^3}$ with $\Omega_{\Lambda 0} = 1 -\Omega_{m0} = 0.6847$ and $H_0 = 67.36$ km s$^{-1}$ Mpc$^{-1}$, we find an expectation value $\vert Y_B\vert = 6.25 \times 10^{-11}$ at $t_{\rm eq}$.

\begin{figure}[t]
	\begin{center}
		\includegraphics[width=7 cm]{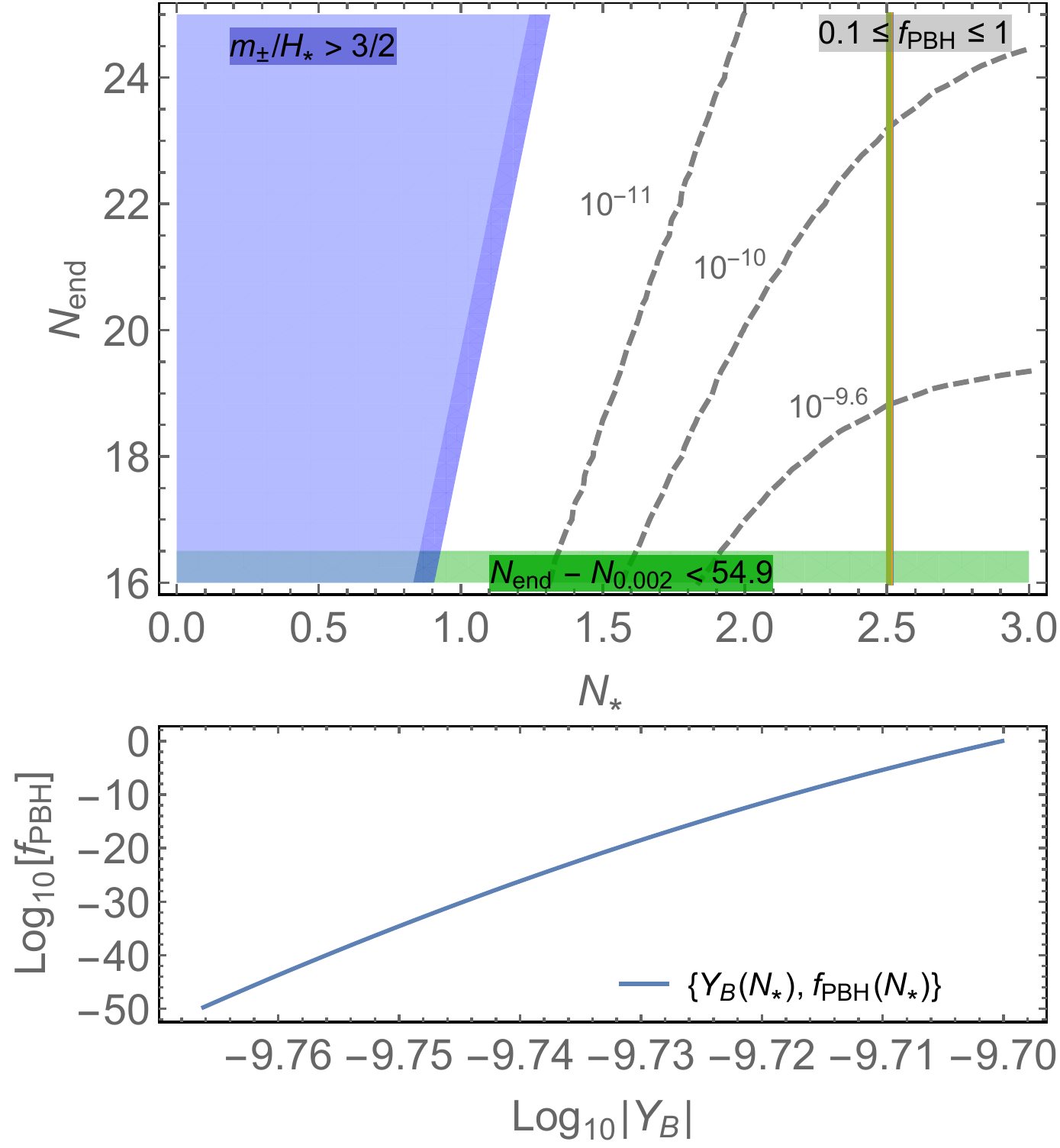}
	\end{center}
	\caption{\label{fig.Nend_Ns}(Upper panel-) Contours of the final baryon asymmetry $\vert Y_B \vert$ in radiation domination with $\delta = -3.15$, $m_\sigma = H_\ast/2$, $\Lambda = 0.3 M_P$, where $H_\ast = 2.37 \times 10^{13}$ GeV and $\Gamma_\phi = 10^{13}$ GeV are used. The region $2.506< N_\ast < 2.515$ corresponds to the PBH-to-DM ratio $0.1 < f_{\rm PBH} < 1$ for $\delta = -3.15$ at the pivot scale $k_0 =9.46 \times 10^{13}$ Mpc$^{-1}$.
	(Lower panel-) $f_{\rm PBH}$ and $\vert Y_B \vert$ as functions of $N_\ast$ at $N_{\rm end} = 20$ and $\delta = -3.15$ from $f_{\rm PBH} = 1$ at $N_\ast = 2.515$ to $f_{\rm PBH} = 10^{-50}$ at $N_\ast = 2.3$.
	}
\end{figure}

An example of a  parameter scan for the $\vert Y_B \vert$ given by \eqref{Y_B__final} at the beginning of radiation domination is given in Fig.~\ref{fig.Nend_Ns} with $\delta = -3.15$, $m_\sigma/H_\ast = 0.5$ and $\Lambda/M_P = 0.3$. $Y_B$ is assumed to be a conserved quantity until matter-radiation equality. $Y_B \gtrsim 10^{-10}$ can be reached with $N_\ast > 1.5$ (for $N_{\rm end} < 18$), which translates to $A_{\rm PBH} > 4.4\times 10^{-5}$. Changing $A_{\rm PBH}$ by 1 order of magnitude roughly corresponds to a $0.35$ variation in $N_\ast$.    

The PBH abundance is exponentially sensitive to the peak value $\nu\equiv \Delta/\sigma_0$ in all statistical methods, which means that a tiny change in $A_{\rm PBH}$ will result in a large difference to $\Omega_{\rm PBHeq}$ or $f_{\rm PBH}$. As a result, the condition $f_{\rm PBH} > 0.1$ for PBH to be an important DM contributor, indeed, specifies a very precise parameter space for the USR inflation. 

\begin{figure}
	\centering
		\includegraphics[width=8 cm]{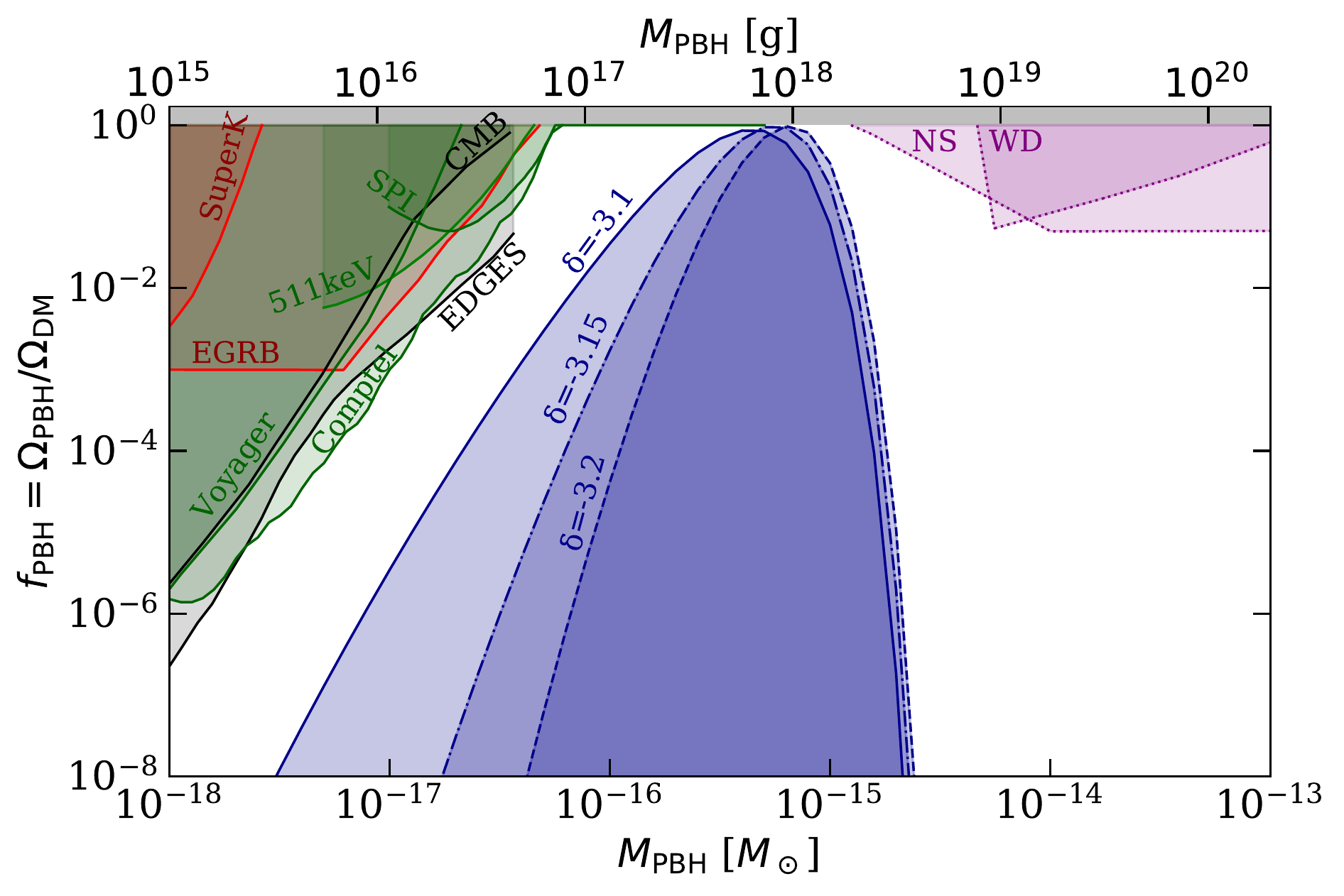}
	\caption{\label{fig.mass_function} The PBH mass function based on the fiducial Press-Schechter statistics at $N_{\rm end}  = 20$ with various choices of $\delta$, where $N_\ast$ is fixed by the condition $f_{\rm PBH} = 1$. The pivot scale $k_0 =9.46 \times 10^{13}$ Mpc$^{-1}$ is used. Existing observational constraints are also shown, using the publicly available code \href{https://github.com/bradkav/PBHbounds}{\underline{PLOTBOUNDS}}.}
\end{figure}

To explore the fiducial parameter space for PBH DM, we adopt the standard Press-Schechter (PS) method \cite{Carr:1975qj} based on the linear density relation $P_\Delta = 16/81 (kR)^4 P_\zeta$ to obtain the mass fraction $\beta_{\rm PS}(M_H) = \textrm{erfc}(\nu_c/\sqrt{2})$ in terms of inflation parameters $\{\delta,N_\ast,N_{\rm end}\}$ with a detailed analytic expression given  in \cite{Wu:2021mwy}.
The mass function $f(M_H)$ based on $\beta_{\rm PS}(M_H)$ with various choices of $\delta$ is displayed in Fig.~\ref{fig.mass_function} (blue shadowed regions). 


We compare the PBH abundance with the existing observational constraints in the literature. The Galactic constraints are displayed in green. They include the bounds from: the local flux of $e^\pm$ measured by {\sc Voyager1} \cite{Boudaud:2018hqb}, the MeV diffuse flux observed by the {\sc INTEGRAL/SPI} detector \cite{Laha:2020ivk}, the 511 keV line in our Galaxy \cite{Laha:2019ssq, DeRocco:2019fjq}, the primary photons detected by the {\sc Comptel} experiment \cite{Coogan:2020tuf}. The red curves refer to the extragalactic constraints, comprising the diffuse neutrino background measured by {\sc Super-Kamiokande} \cite{Dasgupta:2019cae} and the extragalactic background radiation \cite{Carr:2009jm}. The black lines denote the constraints from the energy injection on the cosmic microwave background at recombination \cite{Clark:2016nst, Poulin:2016anj} as well as the evaporating constraints from the 21 cm signal observed by {\sc EDGES} \cite{Clark:2018ghm, Hektor:2018qqw,Mittal:2021egv}. The figure does not include the bounds from the heating of the interstellar medium in dwarf galaxies \cite{Kim:2020ngi}, recently found  in the analysis of \cite{Laha:2020vhg}. The dynamical constraints based on the destruction of white dwarfs and neutrons stars by PBHs are displayed in purple. They are shown with dash-dotted lines, since they are  controversial \cite{Capela:2014qea,Defillon:2014wla}. The interested reader can find a comprehensive discussion on these constraints and future prospects in \cite{Ali-Haimoud:2019khd, Carr:2020gox, Green:2020jor}.

The parameter space for $0.1 < f_{\rm PBH} < 1$ based on $\beta_{\rm PS}$ with $\delta = -3.15$ is given in Fig.\ref{fig.YB_Ns}. $f_{\rm PBH}$ is invariant with respect to $N_{\rm end}$. The difference in $N_\ast$ for $ f_{\rm PBH} = 1$ and $ f_{\rm PBH} = 0.1$ is $\mathcal{O}(10^{-2})$. 
The mass function computed by the peak statistics \cite{Bardeen:1985tr} with additional spatial constraints \cite{Wu:2020ilx,Young:2014ana} shows a $10^{-2}$ difference in $N_\ast$. The uncertainty of $N_\ast$ due to nonlinear effect between $\Delta$ and $\zeta$ could be as large as $\mathcal{O}(10^{-1})$ \cite{Wu:2021mwy}.
$N_{\rm end} < 16.5$ is excluded by the minimal $e$-fold number for the pivot scale $k_0$ to pass the spatial curvature constraint and the finite deviation of scale-invariant $P_\zeta$ in single-field inflation \cite{Akrami:2018odb}. 

Nonperturbative contributions to the curvature perturbation $\zeta$ due to quantum diffusion of the inflaton dynamics can play an important role in the resulting PBH abundance \cite{Biagetti:2018pjj,Ezquiaga:2019ftu,Pattison:2021oen,Figueroa:2020jkf,Biagetti:2021eep}. In general, the non-Gaussian tail of $\zeta$ in the limit of USR inflation ($\delta \rightarrow -3$) can raise the PBH abundance from the standard Gaussian prediction $\beta_{\rm PS}$ by some 10 orders of magnitude so that the real $N_\ast$ for $0.1 < f_{\rm PBH} < 1$ might be shifted toward a smaller value.
However, for the given example in Fig.~\ref{fig.Nend_Ns} with $\delta = -3.15$ and $k_0 =9.46 \times 10^{13}$ Mpc$^{-1}$, we find $\beta_{\rm PS}(N_\ast = 2.5)/\beta_{\rm PS}(N_\ast = 2.2) \gg 10^{100}$ for the viable range of $N_{\rm end}$. This implies that the effect of quantum diffusion seems unlikely to shift the parameter space for $0.1 < f_{\rm PBH} < 1$ to $N_\ast < 2$, leaving the $\mathcal{O}(1)$ ratio $\Omega_{\rm PBHeq}/\Omega_{\rm Beq}$ nearly unchanged in the scenario.

\section{Summary and discussion}
PBHs fostered by the USR transition during inflation can contribute as a significant  DM component. We have shown that such an USR transition of the inflationary background can trigger successful baryogenesis via the AD mechanism. The resulting baryon asymmetry is asymptotically constant toward the long USR duration limit ($N_\ast \gg 1$), allowing the cosmic coincidence ratio $\Omega_{\rm CDM}/\Omega_{\rm B} \sim \mathcal{O}(1)$ to be realized over large $\sim 10^{100}$ uncertainties in the PBH abundance.

The present scenario involves a plethora of  observational tests, especially via many of the future  gravitational wave (GW) experiments. First, the stochastic GW background sourced by enhanced curvature perturbations (at second order) for PBH DM could be measured by space-based laser interferometers \cite{Saito:2008jc,Saito:2009jt,Cai:2018dig,Clesse:2018ogk,Bartolo:2018evs,Domenech:2021ztg}, where the maximal GW density associated with the onset of the USR transition at the $k_0$ of  interest corresponds to the frequency band $f\sim 10^{-3} - 1$ Hz \cite{Ragavendra:2020sop}.
Second, the high scale inflation preferred by USR baryogenesis implies an observable tensor-to-scalar ratio for the next generation CMB measurements close to the current upper bound \cite{Tristram:2020wbi,Akrami:2018odb}.
Third, one of the consequences of AD baryogenesis is that nontopological solitons ($Q$-balls) could have formed due to the fragmentation of scalar condensates during relaxation \cite{Kusenko:1997si}, leading to an enhanced stochastic GW background from second-order curvature perturbations via temporal $Q$-ball domination \cite{White:2021hwi}.
Last but perhaps the most important of all, PBH as a significant component of  DM in the asteroid-mass window could be verified or eliminated by any of the astrophysical projects mentioned at the beginning of this Letter.

\smallbreak
\begin{acknowledgments}
We are grateful to Kalliopi Petraki for helpful discussions and the full support on this project.
E. P. is supported by: the Fermi Research Alliance, LLC under Contract No. DE-AC02-07CH11359 with the U.S. Department of Energy, Office of High Energy Physics; Department of Excellent grant 2018-2022, awarded by the Italian Ministry of Education, University and Research (MIUR); Research grant of the Italo-French University, under Bando Vinci 2020.
Y.-P. W. was supported by the Agence Nationale de la Recherche (ANR) Accueil de Chercheurs de Haut Niveau (ACHN) 2015 grant (“TheIntricateDark” project). 
The project has received funding from the European Union’s Horizon 2020 research and innovation programme under Grant Agreement No. 101002846 (ERC CoG ``CosmoChart'').
\end{acknowledgments}





\begin{thebibliography}{99}

\bibitem{Poulin:2016anj}
V.~Poulin, J.~Lesgourgues and P.~D.~Serpico,
JCAP \textbf{03}, 043 (2017)
[arXiv:1610.10051 [astro-ph.CO]].

\bibitem{Clark:2016nst}
S.~Clark, B.~Dutta, Y.~Gao, L.~E.~Strigari and S.~Watson,
Phys. Rev. D \textbf{95}, no.8, 083006 (2017)
[arXiv:1612.07738 [astro-ph.CO]].

\bibitem{Kim:2020ngi}
H.~Kim,
[arXiv:2007.07739 [hep-ph]].

\bibitem{Boudaud:2018hqb}
M.~Boudaud and M.~Cirelli,
Phys. Rev. Lett. \textbf{122}, no.4, 041104 (2019)
[arXiv:1807.03075 [astro-ph.HE]].

\bibitem{Carr:2009jm}
B.~J.~Carr, K.~Kohri, Y.~Sendouda and J.~Yokoyama,
Phys. Rev. D \textbf{81}, 104019 (2010)
[arXiv:0912.5297 [astro-ph.CO]].

\bibitem{Carr:2016drx}
B.~Carr, F.~Kuhnel and M.~Sandstad,
Phys. Rev. D \textbf{94}, no.8, 083504 (2016)
[arXiv:1607.06077 [astro-ph.CO]].

\bibitem{Ballesteros:2019exr}
G.~Ballesteros, J.~Coronado-Bl\'azquez and D.~Gaggero,
Phys. Lett. B \textbf{808}, 135624 (2020)
[arXiv:1906.10113 [astro-ph.CO]].


\bibitem{DeRocco:2019fjq}
W.~DeRocco and P.~W.~Graham,
Phys. Rev. Lett. \textbf{123}, no.25, 251102 (2019)
[arXiv:1906.07740 [astro-ph.CO]].

\bibitem{Laha:2019ssq}
R.~Laha,
Phys. Rev. Lett. \textbf{123}, no.25, 251101 (2019)
[arXiv:1906.09994 [astro-ph.HE]].

\bibitem{Laha:2020ivk}
R.~Laha, J.~B.~Mu\~noz and T.~R.~Slatyer,
Phys. Rev. D \textbf{101}, no.12, 123514 (2020)
[arXiv:2004.00627 [astro-ph.CO]].

\bibitem{Dasgupta:2019cae}
B.~Dasgupta, R.~Laha and A.~Ray,
Phys. Rev. Lett. \textbf{125}, no.10, 101101 (2020)
[arXiv:1912.01014 [hep-ph]].

\bibitem{Carr:2020gox}
B.~Carr, K.~Kohri, Y.~Sendouda and J.~Yokoyama,
[arXiv:2002.12778 [astro-ph.CO]].

\bibitem{Cang:2020aoo}
J.~Cang, Y.~Gao and Y.~Ma,
JCAP \textbf{05}, 051 (2021)
[arXiv:2011.12244 [astro-ph.CO]].

\bibitem{Cang:2021owu}
J.~Cang, Y.~Gao and Y.~Z.~Ma,
[arXiv:2108.13256 [astro-ph.CO]].

\bibitem{Niikura:2017zjd}
H.~Niikura, M.~Takada, N.~Yasuda, R.~H.~Lupton, T.~Sumi, S.~More, T.~Kurita, S.~Sugiyama, A.~More and M.~Oguri, \textit{et al.}
Nature Astron. \textbf{3}, no.6, 524-534 (2019)
[arXiv:1701.02151 [astro-ph.CO]].

\bibitem{Niikura:2019kqi}
H.~Niikura, M.~Takada, S.~Yokoyama, T.~Sumi and S.~Masaki,
Phys. Rev. D \textbf{99}, no.8, 083503 (2019)
[arXiv:1901.07120 [astro-ph.CO]].

\bibitem{Montero-Camacho:2019jte}
P.~Montero-Camacho, X.~Fang, G.~Vasquez, M.~Silva and C.~M.~Hirata,
JCAP \textbf{08}, 031 (2019)
[arXiv:1906.05950 [astro-ph.CO]].

\bibitem{Smyth:2019whb}
N.~Smyth, S.~Profumo, S.~English, T.~Jeltema, K.~McKinnon and P.~Guhathakurta,
Phys. Rev. D \textbf{101}, no.6, 063005 (2020)
[arXiv:1910.01285 [astro-ph.CO]].


\bibitem{Nemiroff:1995ak}
R.~J.~Nemiroff and A.~Gould,
Astrophys. J. Lett. \textbf{452}, L111 (1995)
[arXiv:astro-ph/9505019 [astro-ph]].

\bibitem{Jung:2019fcs}
S.~Jung and T.~Kim,
Phys. Rev. Res. \textbf{2}, no.1, 013113 (2020)
[arXiv:1908.00078 [astro-ph.CO]].

\bibitem{Katz:2018zrn}
A.~Katz, J.~Kopp, S.~Sibiryakov and W.~Xue,
JCAP \textbf{12}, 005 (2018)
[arXiv:1807.11495 [astro-ph.CO]].

\bibitem{Bai:2018bej}
Y.~Bai and N.~Orlofsky,
Phys. Rev. D \textbf{99}, no.12, 123019 (2019)
[arXiv:1812.01427 [astro-ph.HE]].

\bibitem{Coogan:2020tuf}
A.~Coogan, L.~Morrison and S.~Profumo,
Phys. Rev. Lett. \textbf{126}, no.17, 171101 (2021)
[arXiv:2010.04797 [astro-ph.CO]].

\bibitem{Capela:2013yf}
F.~Capela, M.~Pshirkov and P.~Tinyakov,
Phys. Rev. D \textbf{87}, no.12, 123524 (2013)
[arXiv:1301.4984 [astro-ph.CO]].

\bibitem{Graham:2015apa}
P.~W.~Graham, S.~Rajendran and J.~Varela,
Phys. Rev. D \textbf{92}, no.6, 063007 (2015)
[arXiv:1505.04444 [hep-ph]].

\bibitem{Ray:2021mxu}
A.~Ray, R.~Laha, J.~B.~Mu\~noz and R.~Caputo,
[arXiv:2102.06714 [astro-ph.CO]].


\bibitem{Dutta:2020lqc}
B.~Dutta, A.~Kar and L.~E.~Strigari,
JCAP \textbf{03}, 011 (2021)
[arXiv:2010.05977 [astro-ph.HE]].

\bibitem{Franciolini:2021tla}
G.~Franciolini, V.~Baibhav, V.~De Luca, K.~K.~Y.~Ng, K.~W.~K.~Wong, E.~Berti, P.~Pani, A.~Riotto and S.~Vitale,
[arXiv:2105.03349 [gr-qc]].

\bibitem{Carr:2020xqk}
B.~Carr and F.~Kuhnel,
[arXiv:2006.02838 [astro-ph.CO]].

\bibitem{Green:2020jor}
A.~M.~Green and B.~J.~Kavanagh,
[arXiv:2007.10722 [astro-ph.CO]].



\bibitem{Turner:1979bt}
M.~S.~Turner,
Phys. Lett. B \textbf{89}, 155-159 (1979)

\bibitem{Barrow:1990he}
J.~D.~Barrow, E.~J.~Copeland, E.~W.~Kolb and A.~R.~Liddle,
Phys. Rev. D \textbf{43}, 984-994 (1991)
doi:10.1103/PhysRevD.43.984

\bibitem{Majumdar:1995yr}
A.~S.~Majumdar, P.~Das Gupta and R.~P.~Saxena,
Int. J. Mod. Phys. D \textbf{4}, 517-529 (1995)

\bibitem{Baumann:2007yr}
D.~Baumann, P.~J.~Steinhardt and N.~Turok,
[arXiv:hep-th/0703250 [hep-th]].

\bibitem{Bambi:2008hp}
C.~Bambi, A.~D.~Dolgov and A.~A.~Petrov,
JCAP \textbf{09}, 013 (2009)
[arXiv:0806.3440 [astro-ph]].

\bibitem{Hamada:2016jnq}
Y.~Hamada and S.~Iso,
PTEP \textbf{2017}, no.3, 033B02 (2017)
[arXiv:1610.02586 [hep-ph]].

\bibitem{Hooper:2020otu}
D.~Hooper and G.~Krnjaic,
Phys. Rev. D \textbf{103}, no.4, 043504 (2021)
[arXiv:2010.01134 [hep-ph]].

\bibitem{Datta:2020bht}
S.~Datta, A.~Ghosal and R.~Samanta,
JCAP \textbf{08}, 021 (2021)
[arXiv:2012.14981 [hep-ph]].

\bibitem{JyotiDas:2021shi}
S.~Jyoti Das, D.~Mahanta and D.~Borah,
[arXiv:2104.14496 [hep-ph]].

\bibitem{Ambrosone:2021lsx}
A.~Ambrosone, R.~Calabrese, D.~F.~G.~Fiorillo, G.~Miele and S.~Morisi,
[arXiv:2106.11980 [hep-ph]].

\bibitem{Fujita:2014hha}
T.~Fujita, M.~Kawasaki, K.~Harigaya and R.~Matsuda,
Phys. Rev. D \textbf{89}, no.10, 103501 (2014)
[arXiv:1401.1909 [astro-ph.CO]].

\bibitem{Carr:2019hud}
B.~Carr, S.~Clesse and J.~Garc\'\i{}a-Bellido,
Mon. Not. Roy. Astron. Soc. \textbf{501}, no.1, 1426-1439 (2021)
[arXiv:1904.02129 [astro-ph.CO]].

\bibitem{Garcia-Bellido:2019vlf}
J.~Garc\'\i{}a-Bellido, B.~Carr and S.~Clesse,
[arXiv:1904.11482 [astro-ph.CO]].

\bibitem{Flores:2020drq}
M.~M.~Flores and A.~Kusenko,
Phys. Rev. Lett. \textbf{126}, no.4, 041101 (2021)
[arXiv:2008.12456 [astro-ph.CO]].

\bibitem{Petraki:2013wwa}
K.~Petraki and R.~R.~Volkas,
Int. J. Mod. Phys. A \textbf{28}, 1330028 (2013)
[arXiv:1305.4939 [hep-ph]].

\bibitem{Bell:2011tn}
N.~F.~Bell, K.~Petraki, I.~M.~Shoemaker and R.~R.~Volkas,
Phys. Rev. D \textbf{84}, 123505 (2011)
[arXiv:1105.3730 [hep-ph]].

\bibitem{vonHarling:2012yn}
B.~von Harling, K.~Petraki and R.~R.~Volkas,
JCAP \textbf{05}, 021 (2012)
[arXiv:1201.2200 [hep-ph]].




\bibitem{Yokoyama:1998pt} 
J.~Yokoyama,
Phys.\ Rev.\ D {\bf 58}, 083510 (1998)
[astro-ph/9802357].

\bibitem{Saito:2008em} 
R.~Saito, J.~Yokoyama and R.~Nagata,
JCAP {\bf 0806}, 024 (2008)
[arXiv:0804.3470 [astro-ph]].

\bibitem{Garcia-Bellido:2017mdw} 
J.~Garcia-Bellido and E.~Ruiz Morales,
Phys.\ Dark Univ.\  {\bf 18}, 47 (2017)
[arXiv:1702.03901 [astro-ph.CO]].

\bibitem{Kannike:2017bxn}
K.~Kannike, L.~Marzola, M.~Raidal and H.~Veerm\"ae,
JCAP \textbf{09}, 020 (2017)
[arXiv:1705.06225 [astro-ph.CO]].

\bibitem{Germani:2017bcs} 
C.~Germani and T.~Prokopec,
Phys.\ Dark Univ.\  {\bf 18}, 6 (2017)
[arXiv:1706.04226 [astro-ph.CO]].

\bibitem{Motohashi:2017kbs}
H.~Motohashi and W.~Hu,
Phys. Rev. D \textbf{96}, no.6, 063503 (2017)
[arXiv:1706.06784 [astro-ph.CO]].

\bibitem{Ballesteros:2017fsr}
G.~Ballesteros and M.~Taoso,
Phys. Rev. D \textbf{97}, no.2, 023501 (2018)
[arXiv:1709.05565 [hep-ph]].

\bibitem{Cicoli:2018asa} 
M.~Cicoli, V.~A.~Diaz and F.~G.~Pedro,
JCAP {\bf 1806}, no. 06, 034 (2018)
[arXiv:1803.02837 [hep-th]].

\bibitem{Ozsoy:2018flq}
O.~\"Ozsoy, S.~Parameswaran, G.~Tasinato and I.~Zavala,
JCAP \textbf{07}, 005 (2018)
[arXiv:1803.07626 [hep-th]].





\bibitem{Biagetti:2018pjj}
M.~Biagetti, G.~Franciolini, A.~Kehagias and A.~Riotto,
JCAP \textbf{07}, 032 (2018)
[arXiv:1804.07124 [astro-ph.CO]].

\bibitem{Atal:2018neu}
V.~Atal and C.~Germani,
Phys. Dark Univ. \textbf{24}, 100275 (2019)
[arXiv:1811.07857 [astro-ph.CO]].


\bibitem{Xu:2019bdp}
W.~T.~Xu, J.~Liu, T.~J.~Gao and Z.~K.~Guo,
Phys. Rev. D \textbf{101}, no.2, 023505 (2020)
[arXiv:1907.05213 [astro-ph.CO]].

\bibitem{Motohashi:2019rhu}
H.~Motohashi, S.~Mukohyama and M.~Oliosi,
JCAP \textbf{03}, 002 (2020)
[arXiv:1910.13235 [gr-qc]].

\bibitem{Bhaumik:2019tvl}
N.~Bhaumik and R.~K.~Jain,
JCAP \textbf{01}, 037 (2020)
[arXiv:1907.04125 [astro-ph.CO]].

\bibitem{Mishra:2019pzq}
S.~S.~Mishra and V.~Sahni,
JCAP \textbf{04}, 007 (2020)
[arXiv:1911.00057 [gr-qc]].

\bibitem{Ballesteros:2020qam}
G.~Ballesteros, J.~Rey, M.~Taoso and A.~Urbano,
JCAP \textbf{07}, 025 (2020)
[arXiv:2001.08220 [astro-ph.CO]].

\bibitem{Ragavendra:2020sop}
H.~V.~Ragavendra, P.~Saha, L.~Sriramkumar and J.~Silk,
[arXiv:2008.12202 [astro-ph.CO]].

\bibitem{Taoso:2021uvl}
M.~Taoso and A.~Urbano,
[arXiv:2102.03610 [astro-ph.CO]].


\bibitem{Leach:2001zf}
S.~M.~Leach, M.~Sasaki, D.~Wands and A.~R.~Liddle,
Phys. Rev. D \textbf{64}, 023512 (2001)
[arXiv:astro-ph/0101406 [astro-ph]].

\bibitem{Leach:2000yw}
S.~M.~Leach and A.~R.~Liddle,
Phys. Rev. D \textbf{63}, 043508 (2001)
[arXiv:astro-ph/0010082 [astro-ph]].

\bibitem{Drees:2019xpp}
M.~Drees and Y.~Xu,
Eur. Phys. J. C \textbf{81}, no.2, 182 (2021)
[arXiv:1905.13581 [hep-ph]].


\bibitem{Tsamis:2003px}
N.~C.~Tsamis and R.~P.~Woodard,
Phys. Rev. D \textbf{69}, 084005 (2004)
[arXiv:astro-ph/0307463 [astro-ph]].

\bibitem{Kinney:2005vj}
W.~H.~Kinney,
Phys. Rev. D \textbf{72}, 023515 (2005)
[arXiv:gr-qc/0503017 [gr-qc]].

\bibitem{Martin:2012pe}
J.~Martin, H.~Motohashi and T.~Suyama,
Phys. Rev. D \textbf{87}, no.2, 023514 (2013)
[arXiv:1211.0083 [astro-ph.CO]].

\bibitem{Motohashi:2014ppa}
H.~Motohashi, A.~A.~Starobinsky and J.~Yokoyama,
JCAP \textbf{09}, 018 (2015)
[arXiv:1411.5021 [astro-ph.CO]].

\bibitem{Anguelova:2017djf}
L.~Anguelova, P.~Suranyi and L.~C.~R.~Wijewardhana,
JCAP \textbf{02}, 004 (2018)
[arXiv:1710.06989 [hep-th]].





\bibitem{Byrnes:2018txb}
C.~T.~Byrnes, P.~S.~Cole and S.~P.~Patil,
JCAP \textbf{06}, 028 (2019)
[arXiv:1811.11158 [astro-ph.CO]].

\bibitem{Cheng:2018qof}
S.~L.~Cheng, W.~Lee and K.~W.~Ng,
Phys. Rev. D \textbf{99}, no.6, 063524 (2019)
[arXiv:1811.10108 [astro-ph.CO]].

\bibitem{Ozsoy:2019lyy}
O.~\"Ozsoy and G.~Tasinato,
JCAP \textbf{04}, 048 (2020)
[arXiv:1912.01061 [astro-ph.CO]].

\bibitem{Liu:2020oqe}
J.~Liu, Z.~K.~Guo and R.~G.~Cai,
Phys. Rev. D \textbf{101}, no.8, 083535 (2020)
[arXiv:2003.02075 [astro-ph.CO]].

\bibitem{Ballesteros:2020sre}
G.~Ballesteros, J.~Rey, M.~Taoso and A.~Urbano,
JCAP \textbf{08}, 043 (2020)
[arXiv:2006.14597 [astro-ph.CO]].

\bibitem{Ng:2021hll}
K.~W.~Ng and Y.~P.~Wu,
[arXiv:2102.05620 [astro-ph.CO]].


\bibitem{Carrilho:2019oqg}
P.~Carrilho, K.~A.~Malik and D.~J.~Mulryne,
Phys. Rev. D \textbf{100}, no.10, 103529 (2019)
[arXiv:1907.05237 [astro-ph.CO]].

\bibitem{Akrami:2018odb}
Y.~Akrami \textit{et al.} [Planck],
[arXiv:1807.06211 [astro-ph.CO]].

\bibitem{Tristram:2020wbi}
M.~Tristram, A.~J.~Banday, K.~M.~G\'orski, R.~Keskitalo, C.~R.~Lawrence, K.~J.~Andersen, R.~B.~Barreiro, J.~Borrill, H.~K.~Eriksen and R.~Fernandez-Cobos, \textit{et al.}
[arXiv:2010.01139 [astro-ph.CO]].

\bibitem{Affleck:1984fy} 
I.~Affleck and M.~Dine,
Nucl.\ Phys.\ B {\bf 249}, 361 (1985).

\bibitem{Linde:1985gh} 
A.~D.~Linde,
Phys.\ Lett.\  {\bf 160B}, 243 (1985).

\bibitem{Dolgov:1991fr} 
A.~D.~Dolgov,
Phys.\ Rept.\  {\bf 222}, 309 (1992).

\bibitem{Dine:1995kz}
M.~Dine, L.~Randall and S.~D.~Thomas,
Nucl. Phys. B \textbf{458}, 291-326 (1996)
[arXiv:hep-ph/9507453 [hep-ph]].

\bibitem{Dine:2003ax}
M.~Dine and A.~Kusenko,
Rev. Mod. Phys. \textbf{76}, 1 (2003)
[arXiv:hep-ph/0303065 [hep-ph]].

\bibitem{Wu:2020pej}
Y.~P.~Wu and K.~Petraki,
JCAP \textbf{01}, 022 (2021)
[arXiv:2008.08549 [hep-ph]].


\bibitem{Wang:2019gbi}
L.~T.~Wang and Z.~Z.~Xianyu,
JHEP \textbf{02}, 044 (2020)
[arXiv:1910.12876 [hep-ph]].

\bibitem{Wang:2020ioa}
L.~T.~Wang and Z.~Z.~Xianyu,
JHEP \textbf{11}, 082 (2020)
[arXiv:2004.02887 [hep-ph]].

\bibitem{Bodas:2020yho}
A.~Bodas, S.~Kumar and R.~Sundrum,
JHEP \textbf{02}, 079 (2021)
[arXiv:2010.04727 [hep-ph]].


\bibitem{Wu:2021mwy}
Y.~P.~Wu, E.~Pinetti, K.~Petraki and J.~Silk,
[arXiv:2109.00118 [hep-ph]].
	
\bibitem{Antoniadis:2011ib}
I.~Antoniadis, P.~O.~Mazur and E.~Mottola,
JCAP \textbf{09}, 024 (2012)
[arXiv:1103.4164 [gr-qc]].
	
	

	
\bibitem{Bardeen:1985tr} 
J.~M.~Bardeen, J.~R.~Bond, N.~Kaiser and A.~S.~Szalay,
Astrophys.\ J.\  {\bf 304}, 15 (1986).

\bibitem{Green:2004wb} 
A.~M.~Green, A.~R.~Liddle, K.~A.~Malik and M.~Sasaki,
Phys.\ Rev.\ D {\bf 70}, 041502 (2004)
[astro-ph/0403181].

\bibitem{Young:2014ana}
S.~Young, C.~T.~Byrnes and M.~Sasaki,
JCAP \textbf{07}, 045 (2014)
[arXiv:1405.7023 [gr-qc]].

\bibitem{Suyama:2019npc} 
T.~Suyama and S.~Yokoyama,
arXiv:1912.04687 [astro-ph.CO].

\bibitem{Wu:2020ilx}
Y.~P.~Wu,
Phys. Dark Univ. \textbf{30}, 100654 (2020)
[arXiv:2005.00441 [astro-ph.CO]].


































\bibitem{Aghanim:2018eyx}
N.~Aghanim \textit{et al.} [Planck],
Astron. Astrophys. \textbf{641}, A6 (2020)
[arXiv:1807.06209 [astro-ph.CO]].

\bibitem{Carr:1975qj} 
B.~J.~Carr,
Astrophys.\ J.\  {\bf 201}, 1 (1975).



\bibitem{Capela:2014qea}
F.~Capela, M.~Pshirkov and P.~Tinyakov,
[arXiv:1402.4671 [astro-ph.CO]].




\bibitem{Laha:2020vhg}
R.~Laha, P.~Lu and V.~Takhistov,
[arXiv:2009.11837 [astro-ph.CO]].

\bibitem{Clark:2018ghm}
S.~Clark, B.~Dutta, Y.~Gao, Y.~Z.~Ma and L.~E.~Strigari,
Phys. Rev. D \textbf{98}, no.4, 043006 (2018)
[arXiv:1803.09390 [astro-ph.HE]].

\bibitem{Hektor:2018qqw}
A.~Hektor, G.~H\"utsi, L.~Marzola, M.~Raidal, V.~Vaskonen and H.~Veerm\"ae,
Phys. Rev. D \textbf{98}, no.2, 023503 (2018)
[arXiv:1803.09697 [astro-ph.CO]].

\bibitem{Mittal:2021egv}
S.~Mittal, A.~Ray, G.~Kulkarni and B.~Dasgupta,
[arXiv:2107.02190 [astro-ph.CO]].

\bibitem{Defillon:2014wla}
G.~Defillon, E.~Granet, P.~Tinyakov and M.~H.~G.~Tytgat,
Phys. Rev. D \textbf{90}, no.10, 103522 (2014)
[arXiv:1409.0469 [gr-qc]].

\bibitem{Ali-Haimoud:2019khd}
A.~Kashlinsky, Y.~Ali-Haimoud, S.~Clesse, J.~Garcia-Bellido, L.~Wyrzykowski, A.~Achucarro, L.~Amendola, J.~Annis, A.~Arbey and R.~G.~Arendt, \textit{et al.}
[arXiv:1903.04424 [astro-ph.CO]].












\bibitem{Ezquiaga:2019ftu}
J.~M.~Ezquiaga, J.~Garc\'\i{}a-Bellido and V.~Vennin,
JCAP \textbf{03}, 029 (2020)
doi:10.1088/1475-7516/2020/03/029
[arXiv:1912.05399 [astro-ph.CO]].

\bibitem{Figueroa:2020jkf}
D.~G.~Figueroa, S.~Raatikainen, S.~Rasanen and E.~Tomberg,
[arXiv:2012.06551 [astro-ph.CO]].

\bibitem{Pattison:2021oen}
C.~Pattison, V.~Vennin, D.~Wands and H.~Assadullahi,
[arXiv:2101.05741 [astro-ph.CO]].

\bibitem{Biagetti:2021eep}
M.~Biagetti, V.~De Luca, G.~Franciolini, A.~Kehagias and A.~Riotto,
[arXiv:2105.07810 [astro-ph.CO]].

\bibitem{Saito:2008jc}
R.~Saito and J.~Yokoyama,
Phys. Rev. Lett. \textbf{102}, 161101 (2009)
[erratum: Phys. Rev. Lett. \textbf{107}, 069901 (2011)]
[arXiv:0812.4339 [astro-ph]].

\bibitem{Saito:2009jt}
R.~Saito and J.~Yokoyama,
Prog. Theor. Phys. \textbf{123}, 867-886 (2010)
[erratum: Prog. Theor. Phys. \textbf{126}, 351-352 (2011)]
[arXiv:0912.5317 [astro-ph.CO]].



\bibitem{Cai:2018dig}
R.~g.~Cai, S.~Pi and M.~Sasaki,
Phys. Rev. Lett. \textbf{122}, no.20, 201101 (2019)
[arXiv:1810.11000 [astro-ph.CO]].

\bibitem{Bartolo:2018evs}
N.~Bartolo, V.~De Luca, G.~Franciolini, A.~Lewis, M.~Peloso and A.~Riotto,
Phys. Rev. Lett. \textbf{122}, no.21, 211301 (2019)
[arXiv:1810.12218 [astro-ph.CO]].

\bibitem{Clesse:2018ogk}
S.~Clesse, J.~Garc\'\i{}a-Bellido and S.~Orani,
[arXiv:1812.11011 [astro-ph.CO]].

\bibitem{Domenech:2021ztg}
G.~Dom\`enech,
[arXiv:2109.01398 [gr-qc]].

\bibitem{Kusenko:1997si}
A.~Kusenko and M.~E.~Shaposhnikov,
Phys. Lett. B \textbf{418}, 46-54 (1998)
[arXiv:hep-ph/9709492 [hep-ph]].



\bibitem{White:2021hwi}
G.~White, L.~Pearce, D.~Vagie and A.~Kusenko,
[arXiv:2105.11655 [hep-ph]].

\end{thebibliography}

\end{document}